\documentclass[prl,onecolumn,graphicx,amssymb,floatfix]{revtex4}
\usepackage{graphicx}
\usepackage{placeins}

\begin{document}

\title{Asking  photons where have they been}
\author{A. Danan, D. Farfurnik,  S. Bar-Ad, and L. Vaidman}
\affiliation{ Raymond and Beverly Sackler School of Physics and Astronomy\\
 Tel-Aviv University, Tel-Aviv 69978, Israel}

\begin{abstract}
We present surprising experimental evidence regarding the past of photons passing through an interferometer. The information about the positions through
which the photons pass in the interferometer is retrieved from modulations of the detected signal at the vibration frequencies of mirrors the photons bounce off.
From the analysis we conclude that the past of the photons is not represented by continuous trajectories, although a ``common sense'' analysis adopted in various welcher weg measurements,  delayed-choice which-path experiments and counterfactual communication demonstrations yields a single
trajectory. The experimental results have a simple explanation in the framework
of the two-state vector formalism of quantum theory.

\end{abstract}
\maketitle

Quantum mechanics does not provide a clear answer to the question: What was the past of a photon which went through an interferometer \cite{past}?
Various {\it welcher weg} measurements \cite{KZ},  delayed-choice which-path experiments \cite{Asp,Per,Kai} and weak-measurements   of photons in interferometers \cite{resh,Stein,Maz} presented  the past of a photon as a trajectory or a set of trajectories. We have carried out experimental weak measurements of the paths of photons going through a nested Mach-Zehnder interferometer, discussed earlier in another context  \cite{Ho,CFVA}, which show a different picture: the past of a photon is not a set of continuous trajectories. The photons tell us that they have been in the parts of the interferometer through which they could not pass! Our results lead to rejection of a ``common sense'' approach to the past of a quantum particle. On the other hand they have a simple explanation within the framework of the two-state vector formalism of quantum theory \cite{ABL,AV90}.

The experiment is analogous to the following scenario: If our radio plays Bach, we know that the  photons come from a classical music station, but if we hear a traffic report, we know that the photons come from a local radio station. We can deduce where the photons were on the basis of the information they provide.

In our experiment we vibrate various mirrors inside the interferometer at different frequencies. The rotation of a mirror causes a vertical shift of the light beam reflected off that mirror. We measure the position of photons coming out of the interferometer, and Fourier-analyze the output signal. When the vibration frequency of a certain mirror appears in the power spectrum, we conclude that photons have been near that particular mirror. The vertical displacement of the beam due to the rotation of the mirrors is significantly smaller than the width of the beam, and the change in the optical path length is much smaller than the wavelength. This ensures that the disruption of the interference in the experiment is negligible.

We start with a Mach-Zehnder interferometer (MZI) aligned in such a way that every photon ends up in detector $D$, see Fig. 1A. Mirrors $A$ and $B$ vibrate around their horizontal axes at frequencies $f_A$ and  $f_B$ respectively, causing an oscillation of the vertical position of the photons reaching the detector. The detector $D$ is a quad-cell photo-detector, and the measured signal is the difference of the currents generated in its upper and lower cells.  As expected, the power spectrum  shows two equal contributions at frequencies  $f_A$ and  $f_B$.

\begin{figure}[t]
 \begin{center} \includegraphics[width=13.5cm]{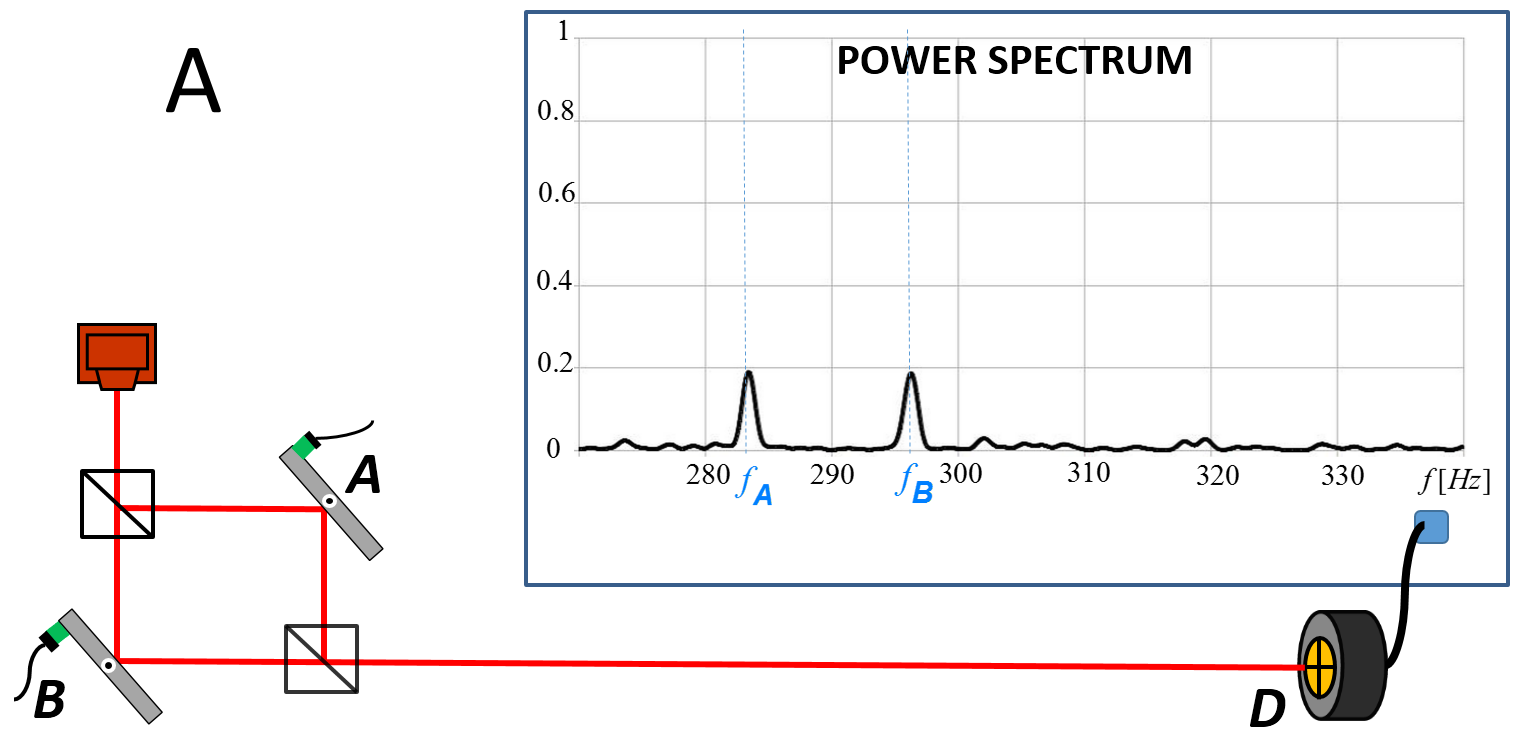}\\ \includegraphics[width=13.5cm]{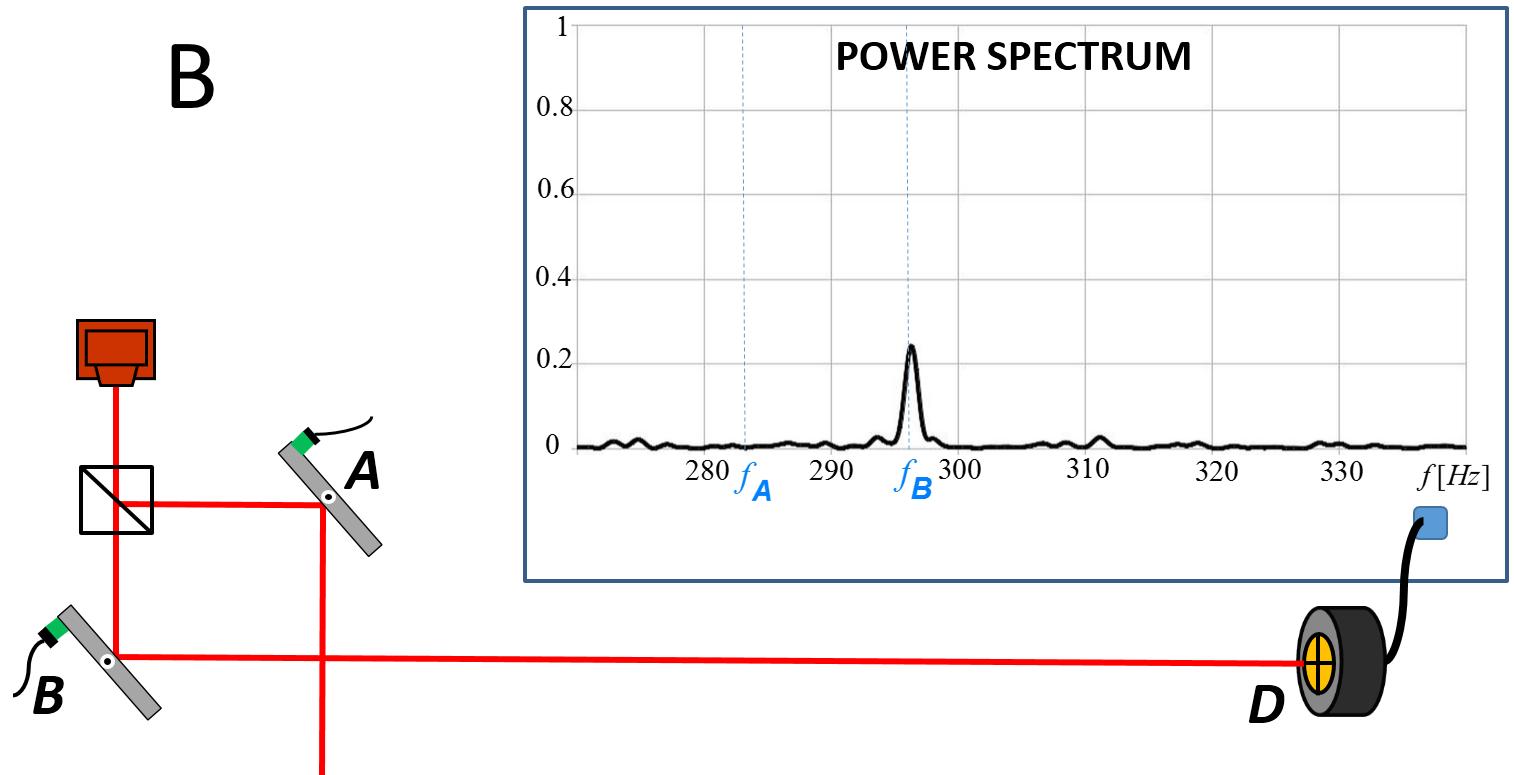}\\
      \caption{(A) Measured power spectrum of the signal from the quad-cell photo-detector shows frequencies of oscillation of internal mirrors $A$ and $B$ of the Mach-Zehnder interferometer. (B) Only the frequency of the mirror $B$ remains in the power spectrum of the signal when the second beam splitter of the interferometer is taken out. } \label{2}\end{center}
\end{figure}

This observation does not answer the question: Did each photon pass through a single arm or through both arms of the interferometer? While the quantum wave of the photon passed through both arms, our measurement does not prove it. The observed current was created by numerous photons, so conceivably different ones were responsible for the two peaks of the power spectrum. Nevertheless, the measurements provide a conclusive evidence that at least some of the observed photons passed near mirrors $A$ and $B$.

We also  measure the signal from the detector when the second beam splitter is removed, transforming the experiment to a which-path measurement, Fig. 1B. Only the signal with frequency  $f_B$ remains.

Next, we build a larger interferometer, Fig.~2A,  in which one third of the beam power goes to the lower arm and two thirds of the beam power pass through the interferometer we just described, in the upper arm. We align the inner (nested) interferometer and the large interferometer in such a way, that again all the photons end up in detector $D$. We vibrate all the mirrors around their horizontal axes with equal (small) amplitudes, each at a different frequency. As expected, the power spectrum shows peaks at all frequencies, and the peaks at frequencies $f_E$ and $f_F$ are higher than the rest.

\begin{figure}[t]
 \begin{center} \includegraphics[width=13cm]{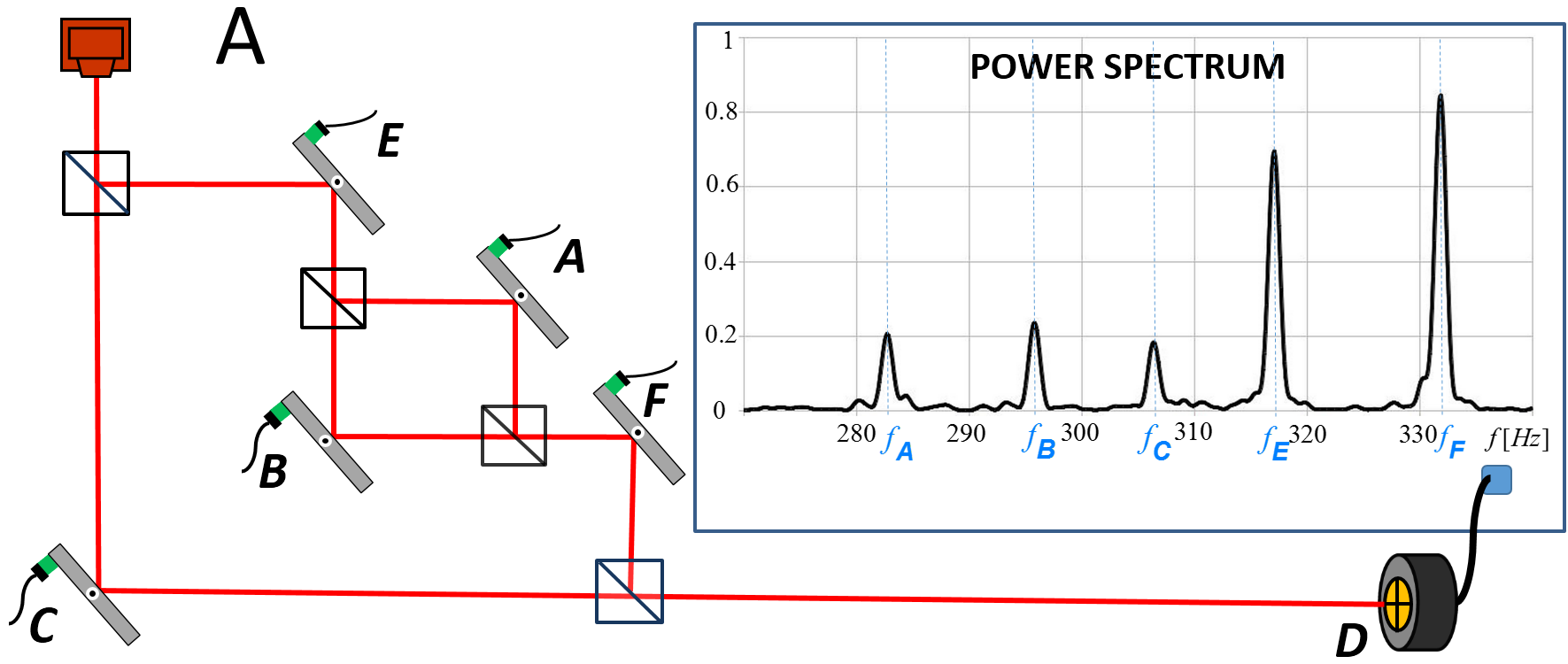}\\\includegraphics[width=13.0cm]{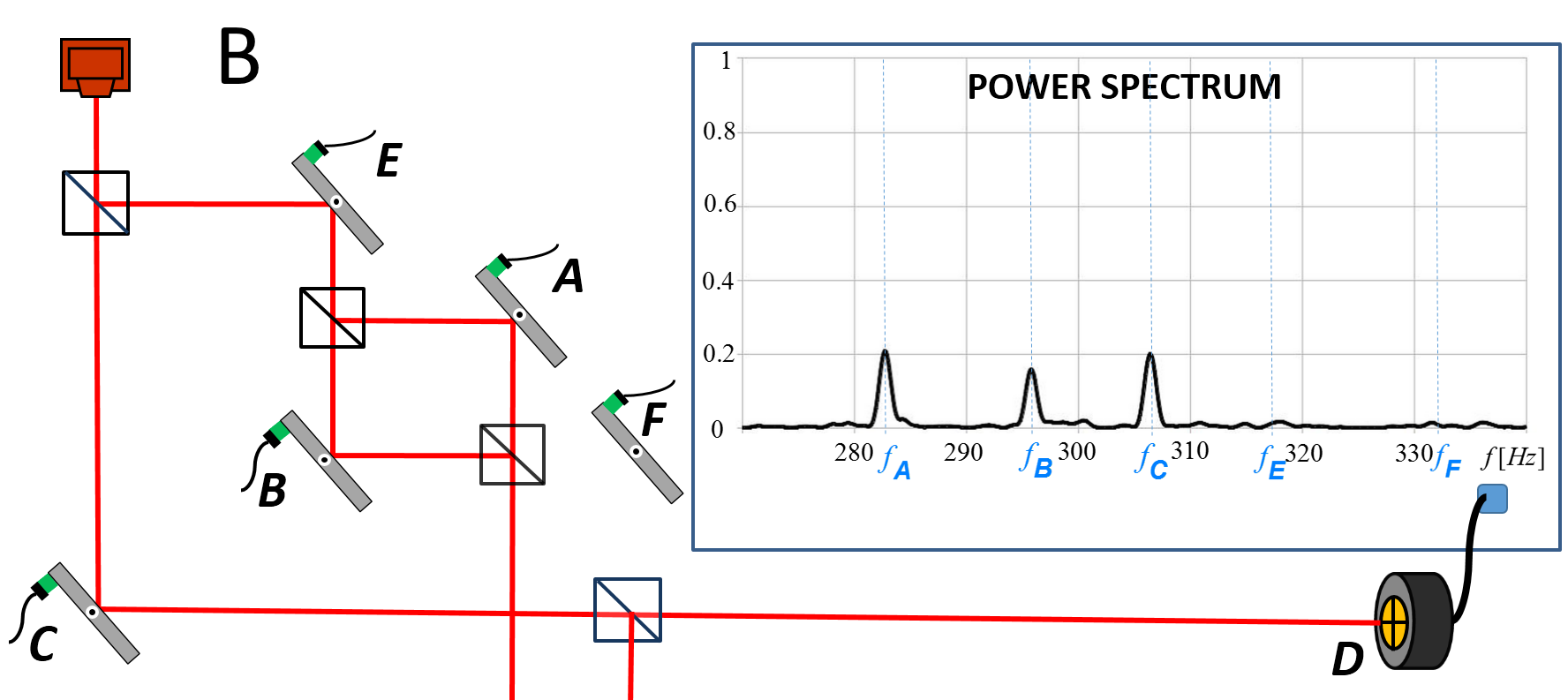}\\\includegraphics [width=13cm]{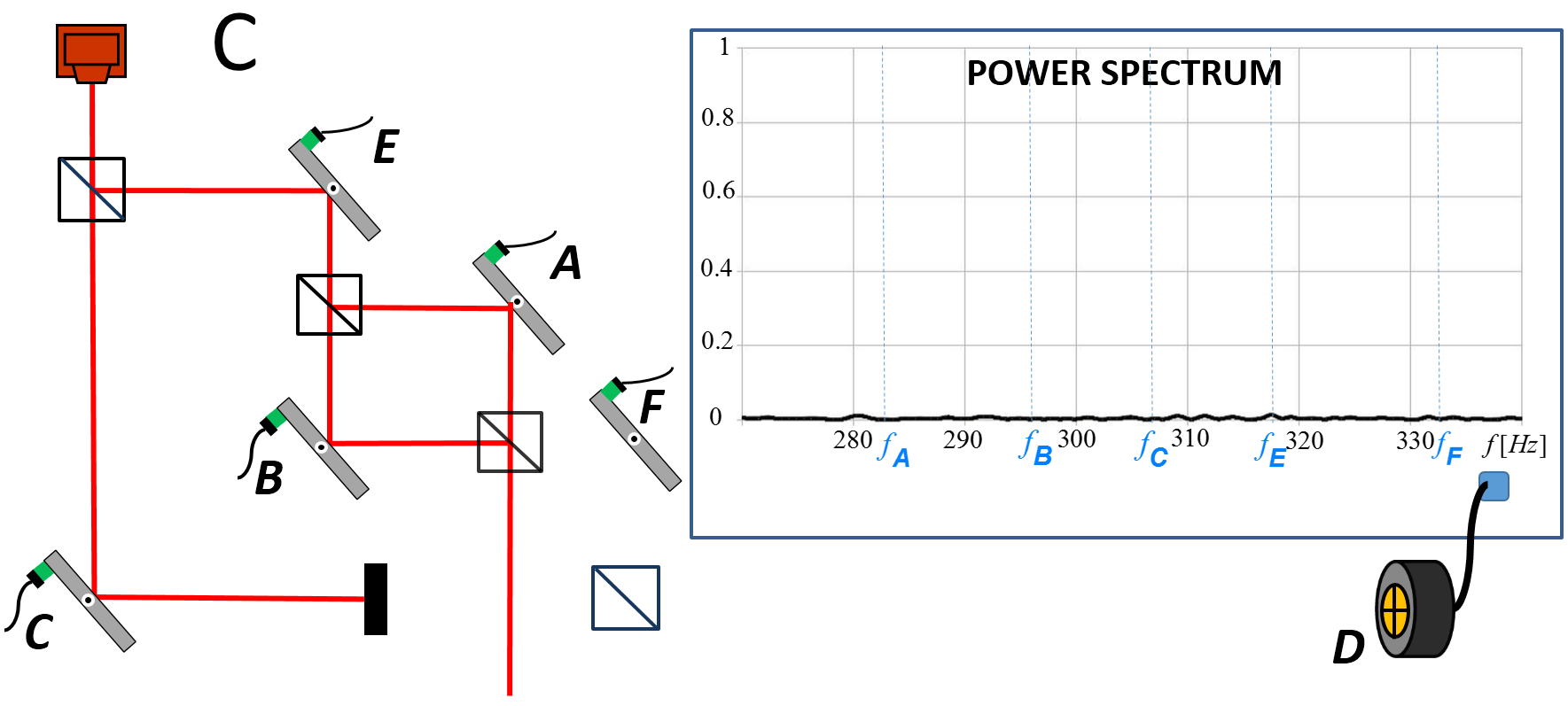}\\
      \caption{(A) Measured power spectrum of the signal from the quad-cell photo detector shows frequencies of oscillation of all internal mirrors of the interferometer. (B) When the inner  interferometer is tuned in such a way that the beam of light passing through it does not reach the photo-detector, the power spectrum of the signal in the photo-detector still shows frequencies of the mirrors of this interferometer. (C) These frequencies (and all other signals) disappear when we, without changing anything in the upper arm, block the lower arm of the large interferometer. } \label{2}\end{center}
\end{figure}

The surprising result is obtained when the interferometer is modified to be a which-path experiment, using the nested MZI as a switch. By slightly shifting mirror $B$ we align the MZI so that there is complete destructive interference of the light propagating towards mirror $F$, see Fig.~2B. In that case observing photons at $D$ supposedly implies that those photons chose the lower arm of the interferometer. We thus expect that the photons reaching detector $D$ have interacted only with mirror $C$, and that the measured signal should therefore show only the frequency $f_C$. Yet the results of the experiment are different: we observe \emph{three} peaks of approximately the same strength: the expected one at frequency $f_C$, and two more peaks at frequencies $f_A$ and $f_B$. The photons tell us that they also interacted with the mirrors of the nested interferometer in the upper arm!  To verify the destructive interference of the light directed towards mirror $F$ we blocked the lower arm, see Fig. 2C. The result was a null signal with no peaks at all. Another surprising feature of the power spectrum of Fig. 2B is the presence of peaks at $f_A$ and $f_B$ and the absence of peaks at $f_E$ and $f_F$. The photons passing through the inner interferometer have to be reflected off the mirrors $E$ and $F$ and thus they are expected to yield even higher peaks at frequencies $f_E$ and $f_F$.

The intuitive picture which allows to foresee and explain these results is given by the two-state vector formulation (TSVF) of quantum mechanics \cite{ABL,AV90}. The observed behavior applies equally if the experiment is performed with single photons \cite{Gra} or with a macroscopic number of photons.
In the TSVF each photon observed by detector $D$ is described by the backward-evolving quantum state $\langle \Phi |$ created at the detector, in addition to the standard, forward-evolving wave function $|\Psi \rangle$ created at the source. The equations of the TSVF imply  that a photon can have a local observable effect only if both the forward- {\it and} backward-evolving quantum waves are non-vanishing at this location.
Fig.~3. shows the forward and the backward evolving states inside the interferometer in the setup of Fig.~2B. It does explain the peculiar features of the power spectrum. The peaks at frequencies $f_A$, $f_B$, and $f_C$ correspond to the fact that near mirrors $A$, $B$, and $C$ both forward evolving and backward evolving quantum states are present.
 The absence of the peak at $f_E$ follows from the absence of the backward evolving wave near $E$ while the absence of the forward evolving wave near $F$ explains why we do not see a peak at $f_F$. This picture also explains the power spectrum in Fig. 1A,B, Fig.~2A. and tells us that blocking mirror $F$ should eliminate peaks at frequencies $f_A$ and $f_B$ \cite{Sup1}.

\begin{figure}[t]
 \begin{center} \includegraphics[width=13.5cm]{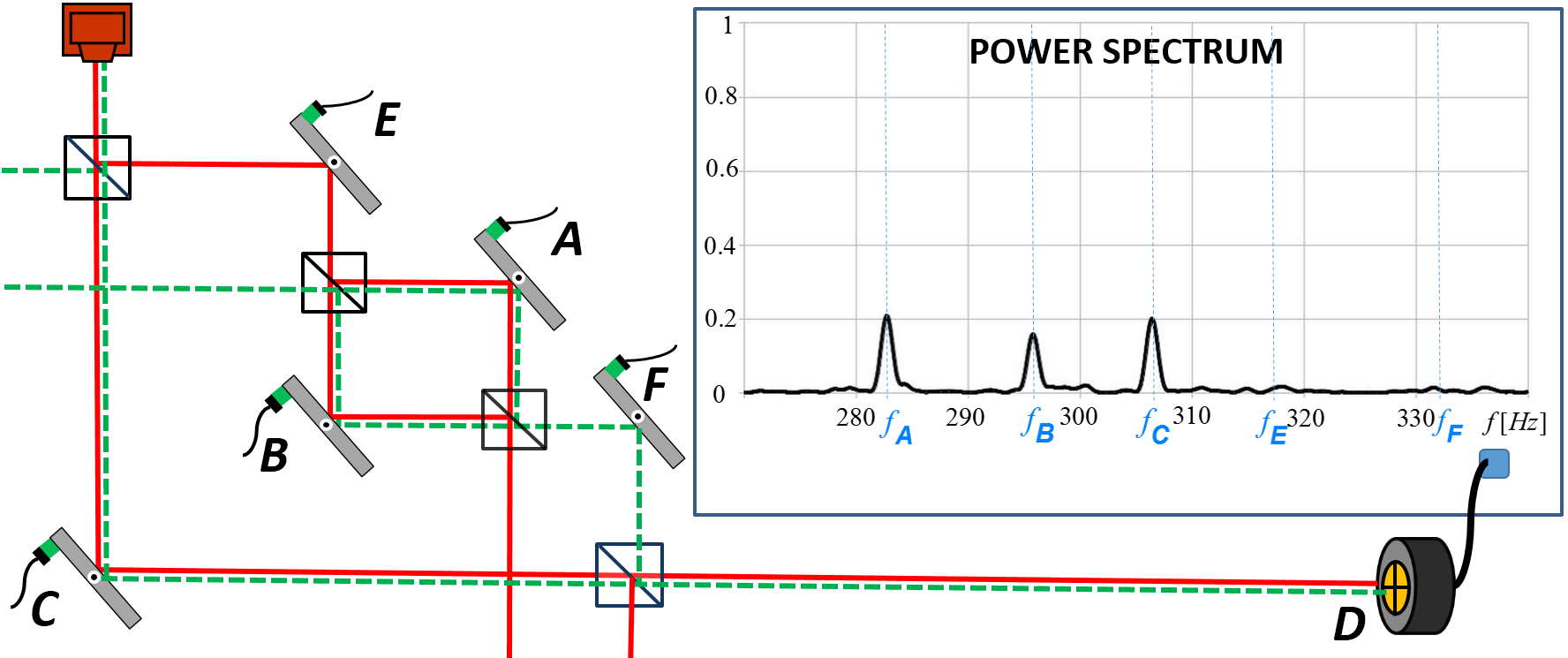}\\
      \caption{The two-state vector description of the photon inside the interferometer includes the standard forward evolving quantum state (red line) and  backward evolving quantum state (green dashed line) of the photon detected by the quad-cell photo-detector. It provides an explanation of the observed power spectrum: frequencies $f_C, f_A$ and $f_B$ are present while $f_E$ and  $f_F$ are not. The photon was present only where both forward and backward quantum wave functions do not vanish. } \label{2}\end{center}
\end{figure}

 Let us describe the experiment depicted in Fig. 2B in the framework of the TSVF in more detail. The two-state vector of the photon at the moment that its partial wave packets bounce off the mirrors $A$, $B$ and $C$  is
 \begin{equation}\label{tsv}
 \nonumber
\langle{\Phi} \vert ~  \vert\Psi\rangle =
{1\over \sqrt 3}\left( \langle A \vert + i\langle B \vert  +   \langle C \vert\right )  ~~ {1\over \sqrt 3}\left( \vert A \rangle +   i\vert B \rangle +  \vert C \rangle \right ) ,
\end{equation}
where we used natural notation: $\vert A \rangle$ is a localized wave packet near mirror $A$, etc. (Note that an equivalent  two-state vector describes the particle in the so called ``Three-box paradox'' \cite{AV91}.)

The main result of the TSVF is that any weak enough coupling to a variable $O$ of a pre- and postselected system  results in an effective coupling to the {\it weak value} of $O$:
  \begin{equation}\label{wv}
O_w \equiv { \langle{\Phi} \vert O \vert\Psi\rangle \over
\langle{\Phi}\vert{\Psi}\rangle }  .
\end{equation}
The rotation of a mirror makes
 the photon itself the measuring device of its projection operator on the location of the mirror. Since the amplitude of the vibration of the mirrors is $\sim  1.5 \cdot 10^{-7} rad$ while the quantum uncertainty  of the direction of the photon is $\sim 3.7 \cdot 10^{-4} rad$, this is a {\it weak measurement} of the projection.
 The pointer variable is the transverse momentum of the photon which is transformed to a spatial shift at the detector.
Using  (\ref{tsv}) and (\ref{wv})  we calculate the weak values of the projection operators at all mirrors  $A, ~B$ and  $C$:
\begin{equation}\label{wvs}
({\rm \bf P}_A)_w =({\rm \bf P}_C)_w =1,~~({\rm \bf P}_B)_w =-1,~~({\rm \bf P}_E)_w =({\rm \bf P}_F)_w =0.
\end{equation}
 This explains the equal peaks at frequencies $f_A, f_B$, and $f_C$ and the absence of peaks at $f_E$ and at $f_F$ in Fig. 3. The peak at $f_B$ is the same as at $f_A$ and $f_C$ because the power spectrum shows only the size of the signal, not its sign. Weak value analysis of the results in other setups presented in Supplement II \cite{Sup2}.

\begin{figure}[b]
 \begin{center} \includegraphics[width=14.5cm]{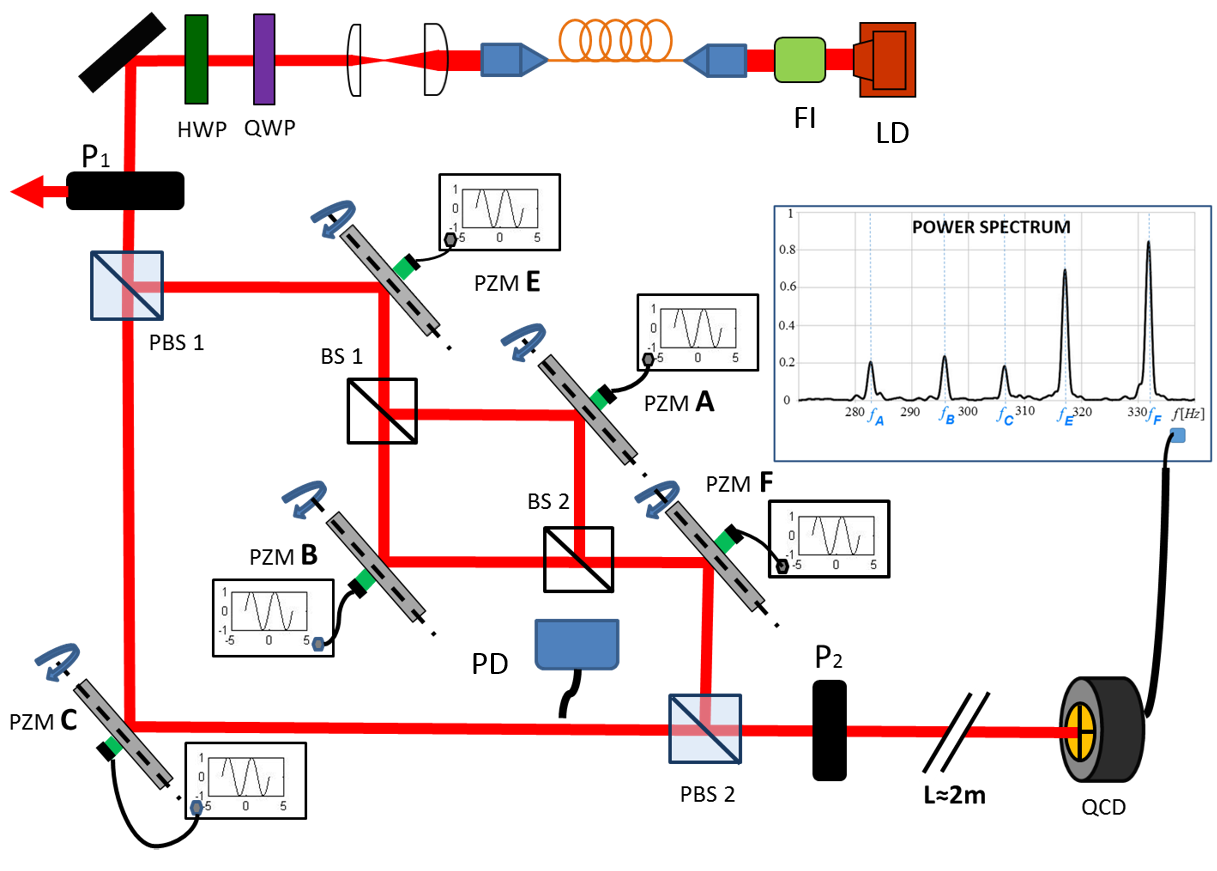}\\
      \caption{Detailed experimental setup.}\end{center}
\end{figure}

Let us describe now the experimental details. The nested MZI inside the upper arm of the large interferometer is constructed from two non-polarizing beam splitters and two mirrors, Fig. 4. The interference in this MZI is controlled by translating mirror $B$ and monitoring the output signal at the left port of $BS2$ using a photo-detector (PD). The measured interference visibility of the inner MZI is about $98.5\%$.  For the external interferometer, 1:2 beam splitters are implemented utilizing polarizers and polarization beam splitters.  The input beam is produced using a continuous wave diode laser (LD) ($λ\sim785nm$ – Thorlabs-L785P090) and a Faraday isolator (FI) which cancels back reflections. Using an attenuator, the intensity of the input beam was reduced by a factor of 3 for the setup with constructive  interference, Fig. 2A, relative to the scheme with destructive interference, Fig. 2B. The beam is spatially filtered and shaped to a $\sim 1mm$ waist Gaussian beam using a single-mode fiber followed by a beam contraction telescope.
A quarter-wave plate, a half-wave plate, and a Glan-Taylor polarizer $P_1$ cancel the birefringence introduced by the fiber, and produce a beam with a linear polarization at angle $54.7^{\circ}$, which leads to an intensity ratio of 2 at the output ports of polarization  beam splitter $PBS_1$.
The beam recombines at a polarization beam splitter $PBS_2$, and is postselected using polarizer $P_2$, also at angle  $54.7^{\circ}$. Mirrors $A, B, C, E$ and $F$ are mounted on piezo-electrically-driven mirror mounts (PZMs), and weak sinusoidal modulations of the tilts around the horizontal axes are used to introduce vertical shifts of the beam reaching the quad-cell photo-detector (QCD) (First Sensor QP50-6-SD2).   Each mirror is seeded with a different sine frequency voltage ($\sim 200mmVpp$) that results in small amplitude oscillations $(\sim 300nrad$) of the reflected beam. At the output of the interferometer we measured the vertical position of the beam using QCD  placed at a distance of $\sim 2m$ from the interferometer. The size of the beam on the detector is $\sim 1.2mm$ and the displacement caused by the mirrors oscillations is about $\sim 600nm$. The QCD signal was sampled using a data acquisition card (NI USB-6008) at a rate of $2.5KHz$ during 1 second intervals. We perform harmonic analysis using MATLAB, and create smoothed power spectra of the signals, Figs.$~1,2$.

Before each data acquisition run, the phase of the nested interferometer was precisely adjusted using a computer-controlled algorithm which brought the intensity at the auxiliary detector PD to minimum (or maximum) by varying the position of mirror $B$ using its PZM. During the one second of each data acquisition run the mirror position was kept fixed, but the phase remained stable. The stability was ensured by preliminary stability analysis \cite{Sup3} and observation of the constant intensity at PD during the run. The phase in the external interferometer was manually tuned before each run - it was not crucial to precisely adjust this phase since it only affected the magnitude of the peaks in the spectrum but not their presence.

Note that the QCD senses photons coming from a wide angle, so the backward evolving state of the detected photon is much wider than what is shown on Fig.$~3$, but this does not change the argument: the backward evolving quantum wave is not reflected off the mirror $E$.

Our interference experiments utilize visible light with about $10^{14}$ photons at each run. Thus, even Maxwell's equations for the classical electromagnetic field should explain the observed phenomena. Indeed, the TSVF adds no new predictions beyond standard quantum mechanics. It just provides a simple intuitive picture of pre- and postselected quantum systems. In the standard framework of  quantum mechanics it is not easy to foresee these effects, but when the question is asked, the calculation is straightforward. The effect is due to a tiny leakage of light in the inner interferometer. Essentially, the same explanation holds also for a classical electromagnetic wave.
 The field entering the interferometer is well characterized  by a Gaussian $\Psi (x,y)={\mathcal A}e^-\frac{x^2+ y^2}{2\Delta^2}$. The beam can reach the quad-cell detector $D$ passing through the three arms $A$, $B$ and $C$. In the setup described in Fig. 2A the fields passing through the different arms all reach  the plane of the quad-cell detector with the same phase, while in Fig. 2B the field passing though arm $B$ gets a relative phase $\pi$. Due to the beam splitters on the way, the amplitudes of the beams passing through each arm are reduced by a factor of 3. Due to the vibration of their angles, the  mirrors cause shifts of the beams in the the $y$ direction such that beam $C$ is shifted by $\delta_C$, beam  $A$ by $\delta_E + \delta_A + \delta_F$, and beam  $B$ by $\delta_E + \delta_B + \delta_F$ .  Thus, in the setup of Fig. 2B, the field on the detector, $\Psi (x,y)$,   is:
\begin{equation}\label{psiPM}
 \nonumber
{\frac{\mathcal A}{3}} (e^{-\frac{x^2+ (y-\delta_C)^2}{2\Delta^2}}+  e^{-\frac{x^2+ (y-\delta_E - \delta_A - \delta_F)^2}{2\Delta^2}}
- e^{-\frac{x^2+ (y-\delta_E - \delta_B - \delta_F)^2}{2\Delta^2}}) .
\end{equation}
The difference of the integrals of the intensity over the regions $y>0$ and $y<0$ is our signal:
  \begin{equation}\label{signal}
 \int_{y>0}|\Psi (x,y)|^2 dx dy - \int_{y<0}|\Psi (x,y)|^2 dx dy .
\end{equation}
Since all the shifts are small,  $\delta \ll \Delta$, only the first order in $\delta$ is considered. Substituting (\ref{psiPM}) in (\ref{signal}) we obtain that  the signal is proportional to  $\delta_C +\delta_A -\delta_B$, in agreement with (\ref{wvs}).
 Every $\delta$ has its own frequency and this explains the observed power spectrum of the signal  \cite{Sup4}.

It will be of interest to repeat this experiment in a regime where neither a classical wave evolution description nor a single particle quantum wave description can provide an explanation. One challenging proposal is to perform the same interference experiment with neutrons. A conceptually different approach aiming to find out where were the particles inside an interferometer would be a weak measurement using an external measuring device. Kerr media provides an interaction between photons, and a weak measurement of a projection operator of a single photon is considered feasible \cite{kerr} for amplified weak values, which, however, is not the case in our setup.

In conclusion, we have performed direct measurements which shed new light on the question: Where were the photons passing through an interferometer? The main results are presented in Fig. 2B. The photons themselves tell us where they have been. And the story they tell is surprising. The photons do not always follow continuous trajectories. Some of them have been inside the nested interferometer (otherwise they could not have known the frequencies $f_A$, $f_B$), but they never entered and never left the nested interferometer, since otherwise they could not avoid the imprints of frequencies $f_E$ and $f_F$ of mirrors $E$ and $F$ leading photons into and out of the interferometer. Only the description with both forward and backward evolving quantum states provides a simple and intuitive  picture of pre- and postselected quantum particles.

{\bf Ackowledgements.}
This work has been supported in part by the Binational Science Foundation Grant No. 32/08, the Israel Science Foundation  Grant No. 1125/10 and Sackler Institute of Condense Matter Equipment Grant.

\pagebreak

\centerline{\LARGE  \bf Supplement I}


\vskip .3cm

\begin{figure}[h!]
 \begin{center} \includegraphics[width=11.5cm]{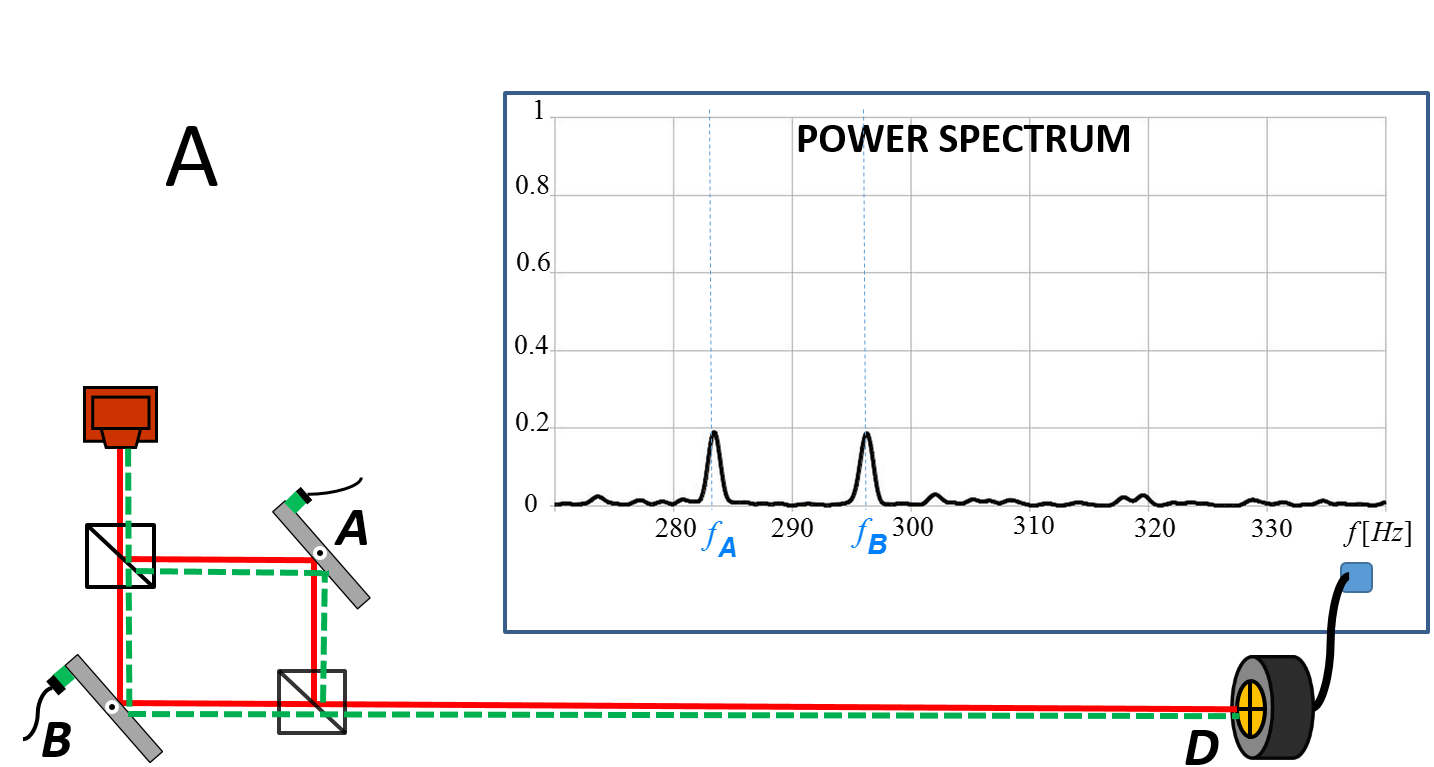}
     \caption{Setup of Fig. 1A. The two-state vector description of the photon inside the interferometer includes the standard forward evolving quantum state (red line) and  backward evolving quantum state (green dashed line) of the photon detected by the quad-cell photo-detector. It provides an explanation of the observed power spectrum: frequencies $f_A$ and $f_B$ appear.}\end{center}
\end{figure}

\begin{figure}[h!]
 \begin{center} \includegraphics[width=11.5cm]{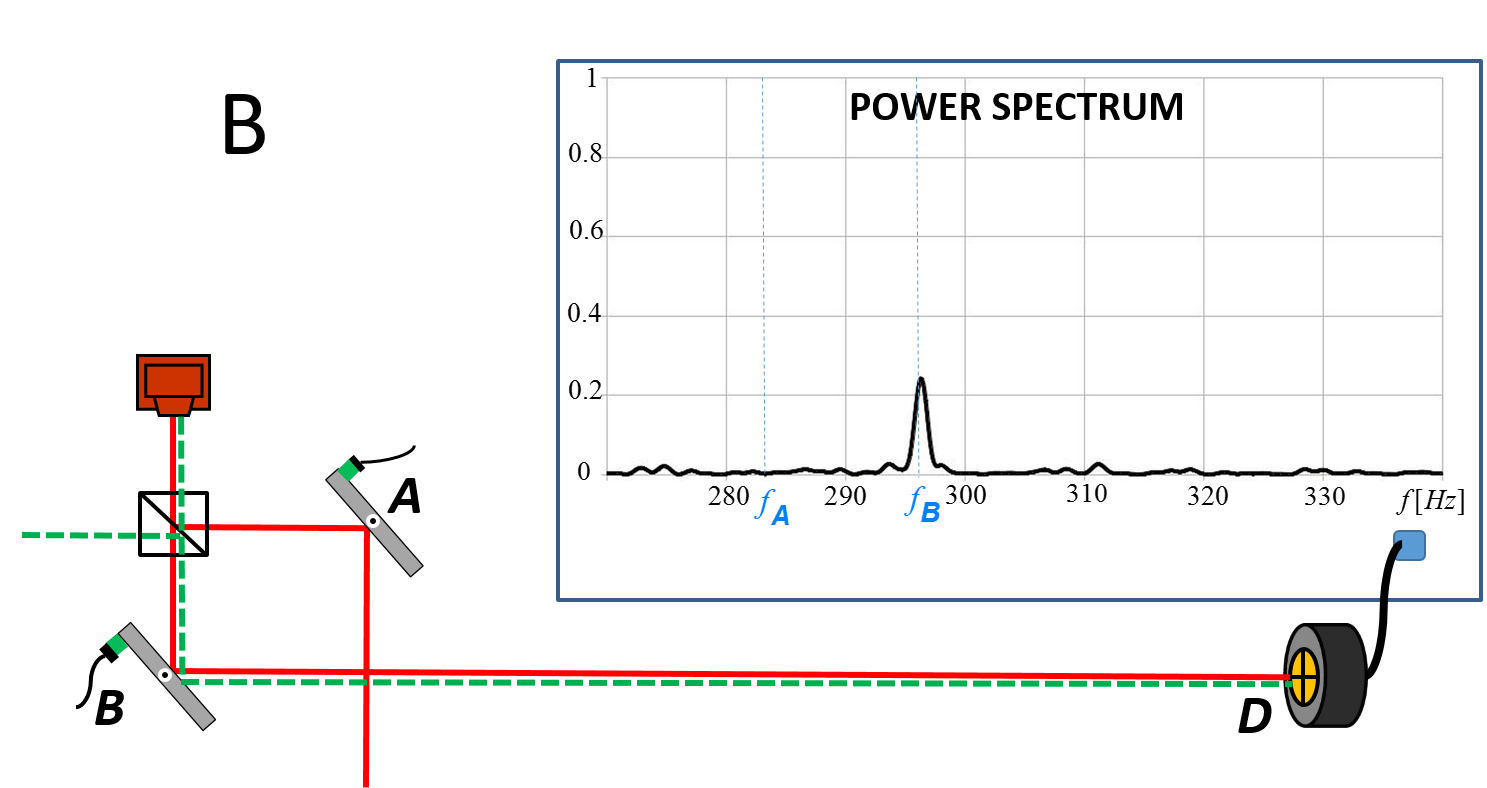} \end{center}
 \caption
      {Setup of Fig. 1B. The  forward evolving  and the  backward evolving quantum states are present together only at mirror $B$: only the peak at frequency  $f_B$  appears. }
\end{figure}

\begin{figure}[h!]
  \includegraphics[width=11cm]{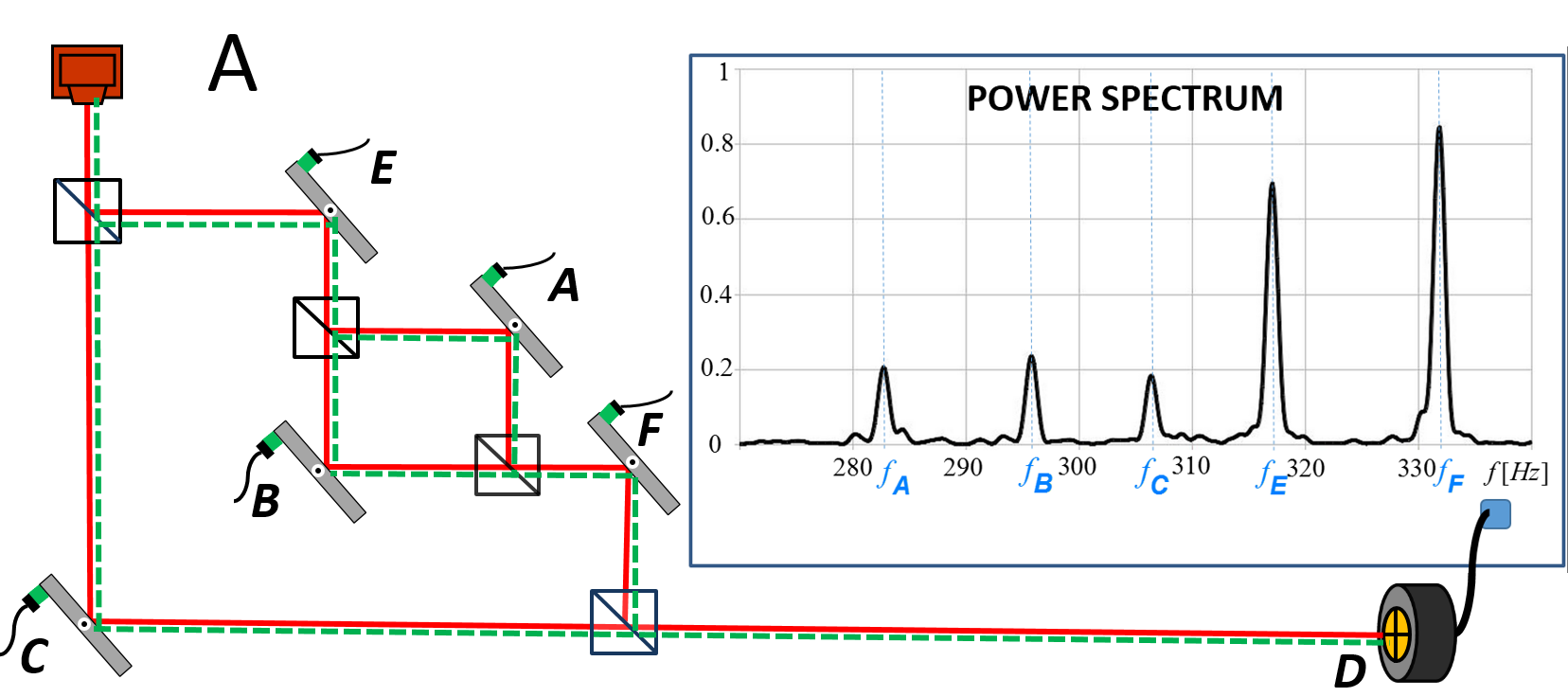}
    \caption
      {Setup of Fig. 2A.The forward and backward evolving states are present together in all mirrors: peaks at all frequencies  appear.}
\end{figure}

\begin{figure}[h!]
 \begin{center} \includegraphics[width=11cm]{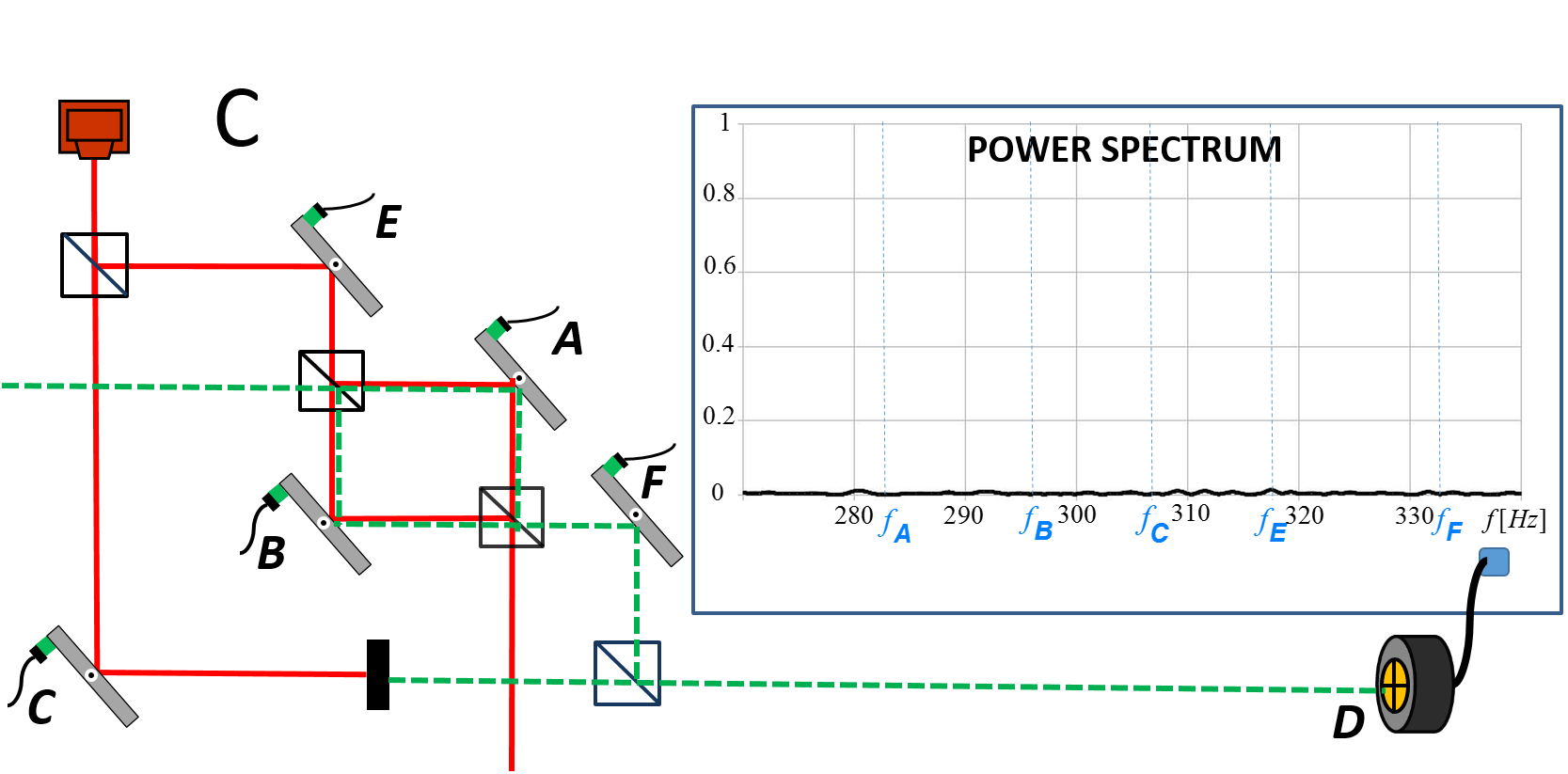}\end{center}
 \caption
   {Setup of Fig. 2C. The  forward evolving  and the  backward evolving quantum states are present together  at mirrors $A$ and $B$, but no peaks are present! The (leaking) photons which do reach the detector are shifted with corresponding frequencies, we just do not get enough photons at the detector to see anything above the noise. }
\end{figure}

\begin{figure}[h!]
 \centerline{\large \bf Additional measurement with  mirror $F$ blocked} \vskip.3cm
 \begin{center} \includegraphics[width=11cm]{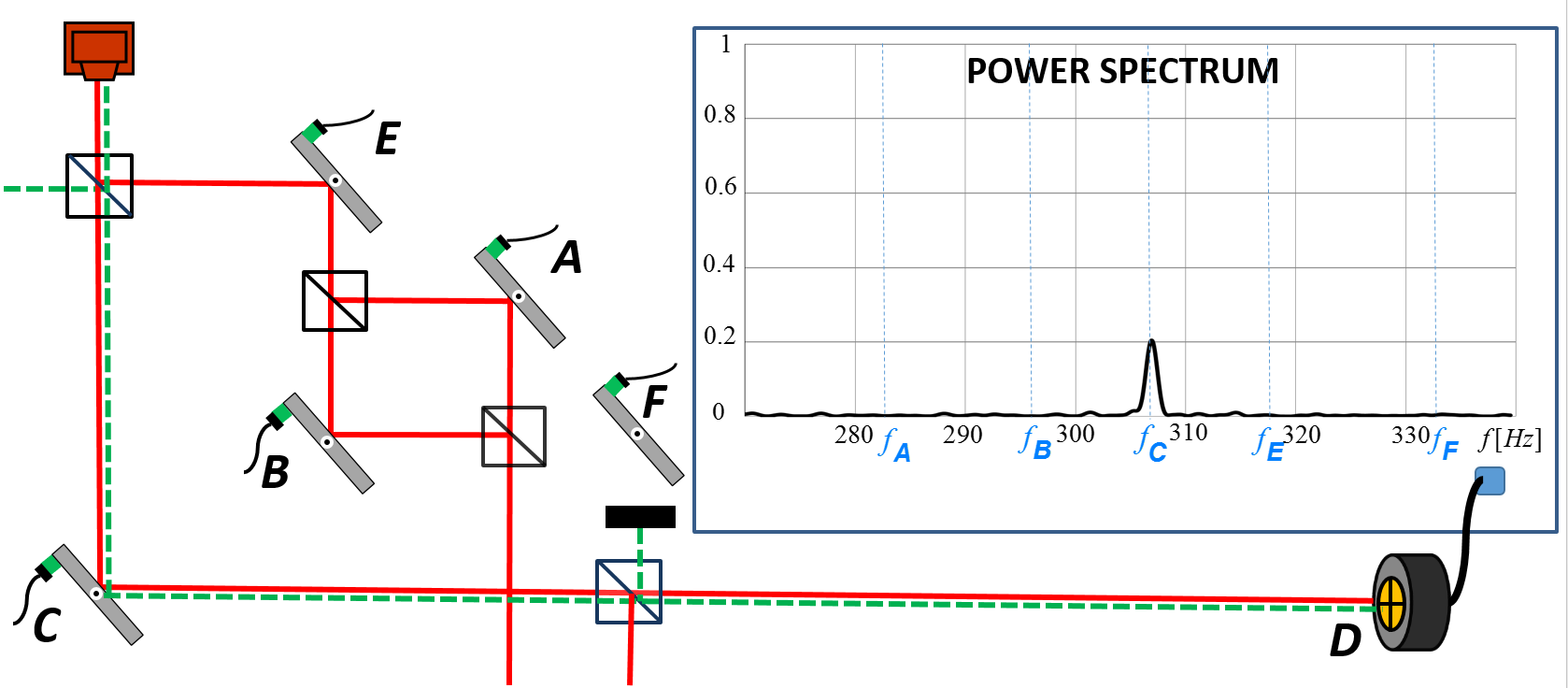}
  \caption{The forward and backward evolving states are present together only at mirror $C$ and thus, the only peak in the power spectrum is at $f_C$. Since according to the naive approach  the  presence of peaks at $f_A$ and $f_B$  in Fig. 2B is very counterintuitive, one might suspect that these peaks may result from some unrelated electronic noise. A block between mirror $F$ and the last beam splitter absorbs the backward evolving wave moving towards mirrors $F$, $A$, $B$, and $E$. In the language of forward wave function only, it absorbs the leakage of the wave from the inner interferometer. Thus theoretically, according to the standard and the TSVF approaches, the block ensures the absence of peaks at the corresponding frequencies. Therefore, this experiment provides a decisive test for the absence of electronic noise  in Fig. 2B.}
\end{center}
\end{figure}

\newpage
\FloatBarrier

\centerline{\LARGE \bf Supplement II}  \vskip .2cm
\centerline{\Large  Weak value analysis of setups Fig. 1A,B and Fig. 2A,C}
 \vskip .2cm


 The main result of the Letter is the presence of the photon near mirrors $A$ and $B$ in the setup of Fig. 2B. as discussed in the Letter itself. However, it is also of interest to understand the presence and the size of the observed peaks in other setups.

 The two state vector of the photon in the setup of Fig. 1A at the intermediate time is:
 \begin{equation}\label{tsv}
 \nonumber
\langle{\Phi} \vert ~  \vert\Psi\rangle =
{1\over \sqrt 2}\left( \langle A \vert + \langle B \vert  \right )  ~~ {1\over \sqrt 2}\left( \vert A \rangle +   \vert B \rangle \right ) ,
\end{equation}
Thus, the weak values of the projections on the locations of the photon at the mirrors are
\begin{equation}\label{wvs}
({\rm \bf P}_A)_w =({\rm \bf P}_B)_w =\frac{1}{2}.
\end{equation}
This explains the equal size peaks at $f_A$ and $f_B$.
\vskip .4cm

The two state vector of the photon in the setup of Fig. 1B at the intermediate time is:
 \begin{equation}\label{tsv}
 \nonumber
\langle{\Phi} \vert ~  \vert\Psi\rangle =
{1\over \sqrt 2}\left( \langle A \vert + \langle B \vert  \right )  ~~    \vert B \rangle  ,
\end{equation}
Thus, the weak value of the projections on the location at the mirrors are
\begin{equation}\label{wvs}
({\rm \bf P}_A)_w =0,~~~~~({\rm \bf P}_B)_w =1.
\end{equation}
This explains why there is only one  peak at  $f_B$. Due to oscillations of mirror  $B$, the shift of the beam is twice as large as in in the setup of Fig. 1A. However, the measured signal is the same since it is also proportional to the total intensity on the quad-cell detector which is half relative to the case of  Fig. 1A.
\vskip .4cm

The two state vector of the photon in the setup of Fig. 2A at the intermediate time is:
\begin{equation}\label{wvs}
\nonumber
\langle{\Phi} \vert ~  \vert\Psi\rangle =
{1\over \sqrt 3}\left( \langle A \vert +  \langle B \vert  +   \langle C \vert\right )  ~~ {1\over \sqrt 3}\left( \vert A \rangle +   \vert B \rangle +  \vert C \rangle \right ) ,
\end{equation}
Thus, the weak value of the projections on the location of the photon at mirrors $A$, $B$, and $C$ are:
\begin{equation}\label{wvs}
({\rm \bf P}_A)_w =({\rm \bf P}_C)_w =({\rm \bf P}_B)_w=\frac{1}{3}.
\end{equation}

At an earlier time, when one of the wave packets is near mirror $E$, the two state vector  is:
\begin{equation}\label{wvs}
\nonumber
\langle{\Phi} \vert ~  \vert\Psi\rangle =
{1\over \sqrt 3}\left(  \sqrt 2 \langle E \vert +    \langle C \vert\right )  ~~ {1\over \sqrt 3}\left( \sqrt 2 \vert E \rangle +    \vert C \rangle \right ) ,
\end{equation}

Thus, the weak value of the projection on the location at the mirror $E$ is:
\begin{equation}\label{wvs}
({\rm \bf P}_E)_w =\frac{2}{3}.
\end{equation}

In a similar way we obtain
\begin{equation}\label{wvs}
({\rm \bf P}_F)_w =\frac{2}{3}.
\end{equation}

These results explain why there are peaks at all frequencies. The peaks at $f_E$ and $f_F$ are four times larger than the peaks at $f_A$,   $f_B$ and $f_C$ because in the definition of the power spectrum, the power is proportional to the square of the signal. The size of the peaks at  $f_A$,   $f_B$ and $f_C$ are the same (up to noise) as in the setup of Fig. 2B in spite of the fact that the weak value (and thus the shift of the beam) is smaller by the factor of 3. This  is because the intensity in the setup of Fig. 2A is 3 times larger. (It had to be 9 times larger but the input intensity was reduced by the factor of 3).

\vskip .4cm

The two-state vector of the photon in the setup of Fig. 2C (Fig. 4 of Supplement 1) at the intermediate time is:
\begin{equation}\label{wvs}
\nonumber
\langle{\Phi} \vert ~  \vert\Psi\rangle =
{1\over \sqrt 3}\left( \langle A \vert +  i\langle B \vert  +   \langle {\it Absorbed} \vert\right )  ~~ {1\over \sqrt 3}\left( \vert A \rangle +   i\vert B \rangle +  \vert C \rangle \right ) ,
\end{equation}
In this situation the weak values are not defined: the post-selected state is orthogonal to the forward evolving state. This means that in the ideal case there are no such pre- and post-selected photons.   This explains why we have seen the null result. (The formalism can deal with  a realistic case in which  the leakage  from the inner interferometer will make the weak values well defined.)
\vskip 1.5cm

\centerline{\LARGE \bf Supplement III}  \vskip .2cm
\centerline{\Large  Stability analysis of the inner interferometer}
 \vskip .1cm

\begin{figure}[h]
 \begin{center} \includegraphics[width=19cm]{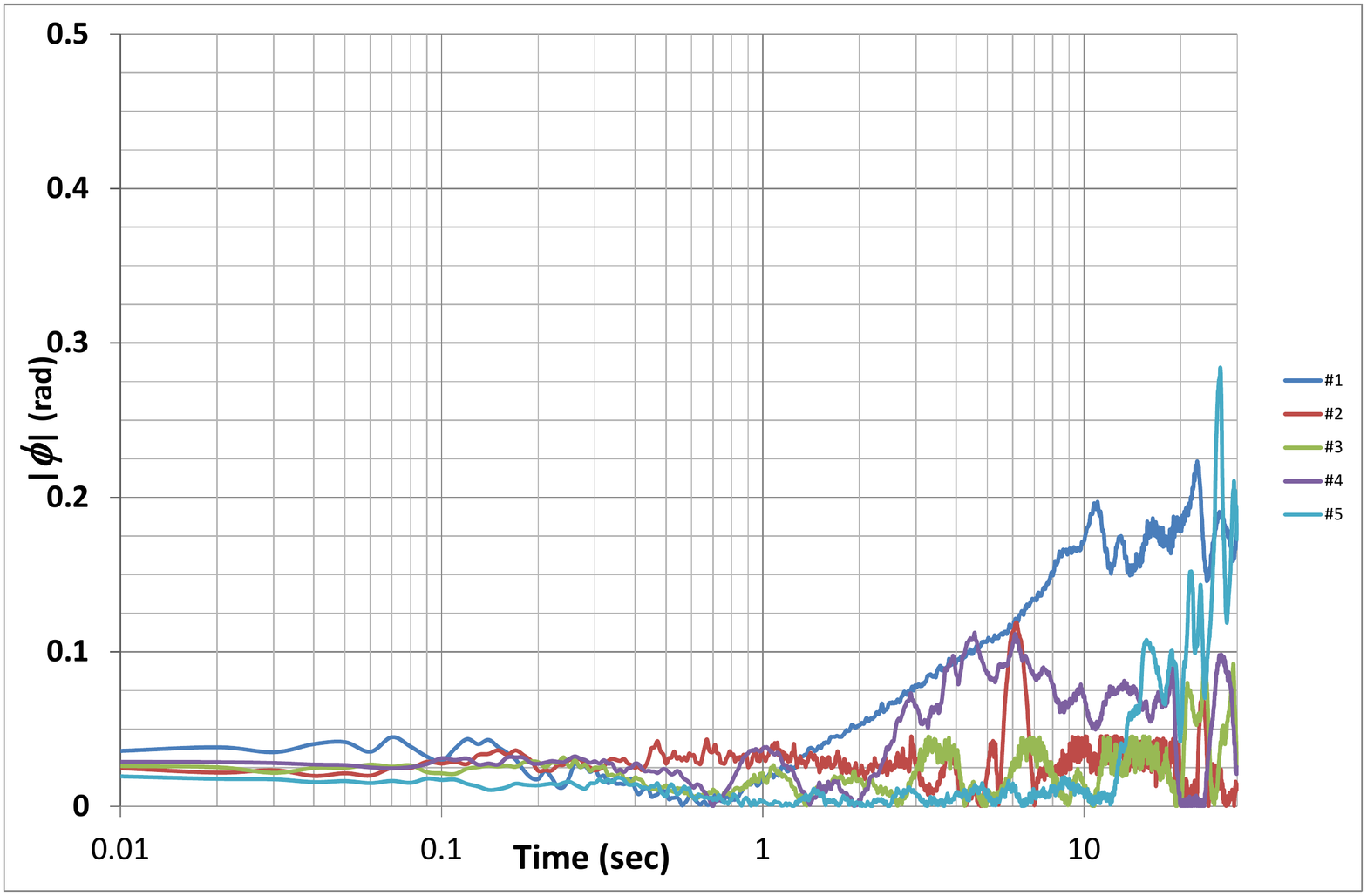}
      \caption{ The absolute value of the phase (calculated based on the intensity measured in the auxiliary detector PD) in five 30-seconds runs. Each run started automatically when the computer-controlled algorithm brought the intensity close to minimum.
      The results show a reliable action of the algorithm that adjust the phase to zero, and the high stability during the first second of passive stabilization - the duration of each run.} \end{center}
\end{figure}

\newpage

\centerline{\LARGE \bf Supplement IV}  \vskip .2cm
\centerline{\Large  Computer calculations of the expected experimental results}
 \vskip .1cm

 Here we present direct calculations of the expected experimental results based on the standard quantum formalism which is the same as in classical electromagnetic waves theory.

 We model the field entering the interferometer  as a Gaussian
 \begin{equation}\label{disp}
 \nonumber
\Psi (x,y)={\mathcal A}e^-\frac{x^2+ y^2}{2\Delta^2} .
\end{equation}

 We model displacement of the beam on the plane of the quad-cell detector due to rotation of a mirror $X$ as
\begin{equation}\label{disp}
 \nonumber
\delta_X=\delta \sin(2\pi f_X t) .
\end{equation}

The parameters corresponding to our experiment are:
\begin{equation}\label{para}
 \Delta=1.2mm,  ~~~\delta=0.6\mu m,
\end{equation}
\begin{equation}\label{para}
 f_A=282 Hz,~ f_B =296 Hz, ~f_C=307Hz, ~f_E=318Hz, ~f_F= 332Hz.
\end{equation}

  We start with the setup described in Fig. 2B.
  The field on the detector is given by Eq. (4) in the Letter:
\begin{equation}\label{psiB}
 \nonumber
\Psi (x,y)={\frac{\mathcal A}{3}} \large(e^{-\frac{x^2+ (y-\delta_C)^2}{2\Delta^2}}+ e^{-\frac{x^2+ (y-\delta_E - \delta_A - \delta_F)^2}{2\Delta^2}}
- e^{-\frac{x^2+ (y-\delta_E - \delta_B - \delta_F)^2}{2\Delta^2}}\large ) .
\end{equation}

Our signal is the difference of the integrals of the intensity over the regions $y>0$ and $y<0$:
  \begin{equation}\label{signal}
 S\equiv\int_{y>0}|\Psi (x,y)|^2 dx dy - \int_{y<0}|\Psi (x,y)|^2 dx dy.
\end{equation}

We have calculated the signal at a rate of $2.5KHz$ during 1 second interval. We performed a harmonic analysis using MATLAB and created smoothed power spectrum of the signal averaging on every 10 points, exactly as in the analysis of the real data.
\begin{figure}[h!]
 \begin{center} \includegraphics[width=12cm]{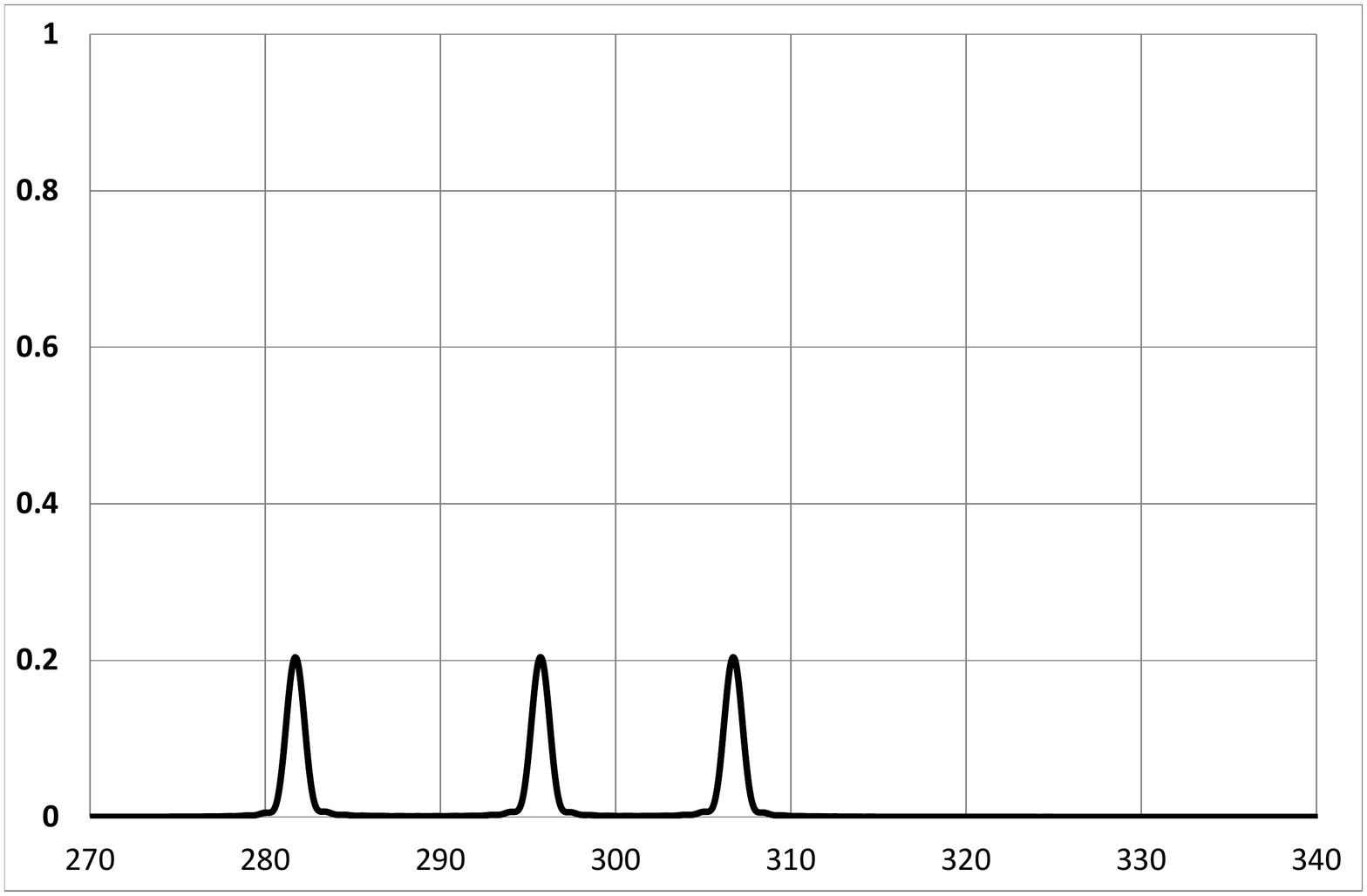}
      \caption{Calculated Power Spectrum of the signal in the setup described in Fig. 2B.
} \label{S2Ai}\end{center}
\end{figure}

\newpage

\FloatBarrier
We repeated this calculation for the setup of Fig. 2C. The field on the detector in this case is:
\begin{equation}\label{psiC}
 \nonumber
\Psi (x,y)={\frac{\mathcal A}{3}} (e^{-\frac{x^2+ (y-\delta_E - \delta_A - \delta_F)^2}{2\Delta^2}}
- e^{-\frac{x^2+ (y-\delta_E - \delta_B - \delta_F)^2}{2\Delta^2}}) .
\end{equation}

\begin{figure}[h]
 \begin{center} \includegraphics[width=10cm]{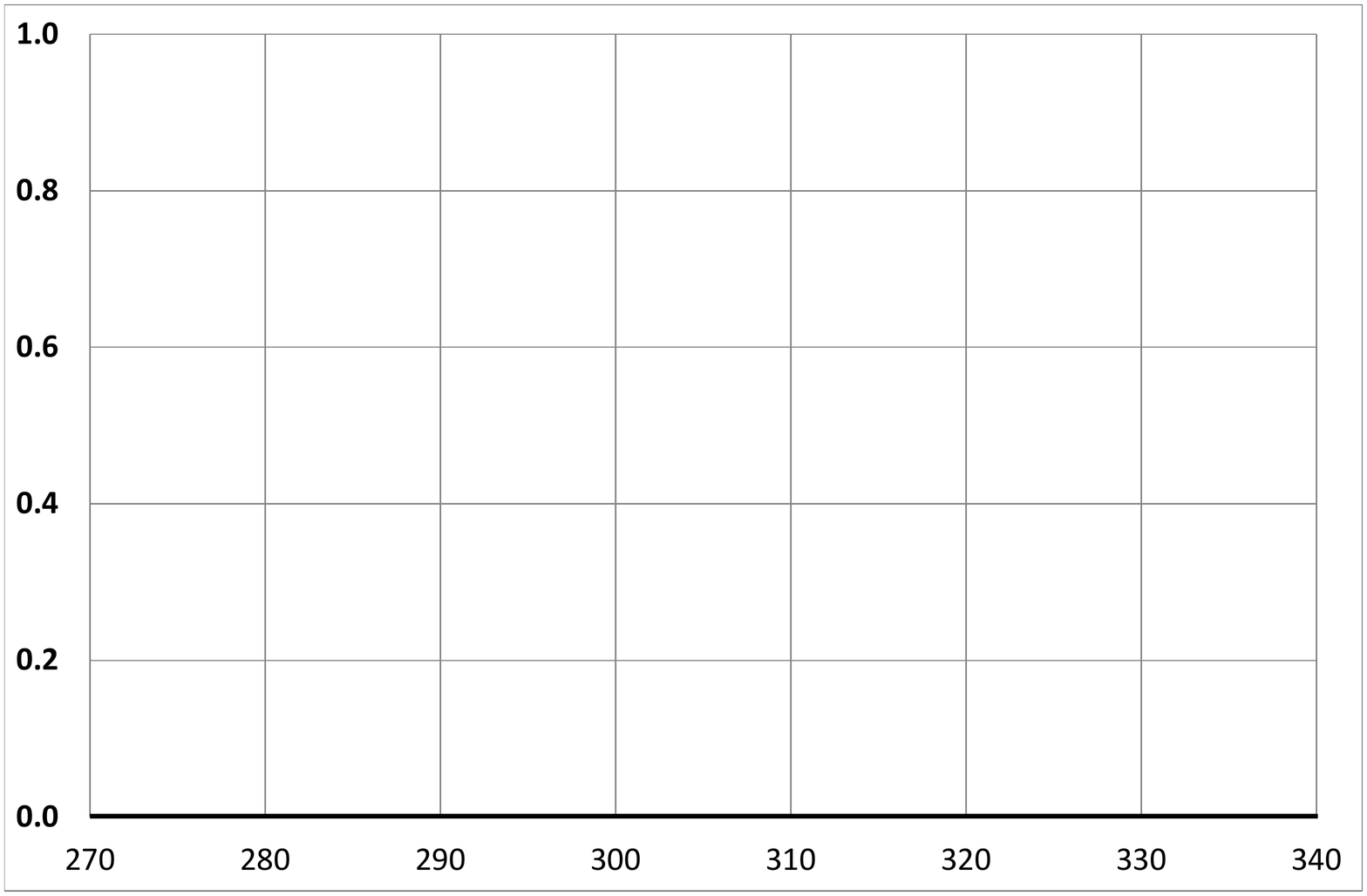}
      \caption{Calculated Power Spectrum of the signal in the setup described in Fig. 2C. } \end{center}
\end{figure}

\FloatBarrier

We repeated this calculation for the setup of Fig. 2A. corresponding to constructive interference. In order to reduce the difference of intensities at the detector between the runs with constructive and destructive interference (a factor of 9) we attenuated the intensity here by a factor of 3. The field in this case is:

\begin{equation}\label{psiB}
 \nonumber
\Psi (x,y)={\frac{\mathcal A}{3\sqrt 3}} (e^{-\frac{x^2+ (y-\delta_C)^2}{2\Delta^2}}+ e^{-\frac{x^2+ (y-\delta_E - \delta_A - \delta_F)^2}{2\Delta^2}}
+ e^{-\frac{x^2+ (y-\delta_E - \delta_B - \delta_F)^2}{2\Delta^2}}) .
\end{equation}

\begin{figure}[h!]
 \begin{center} \includegraphics[width=10cm]{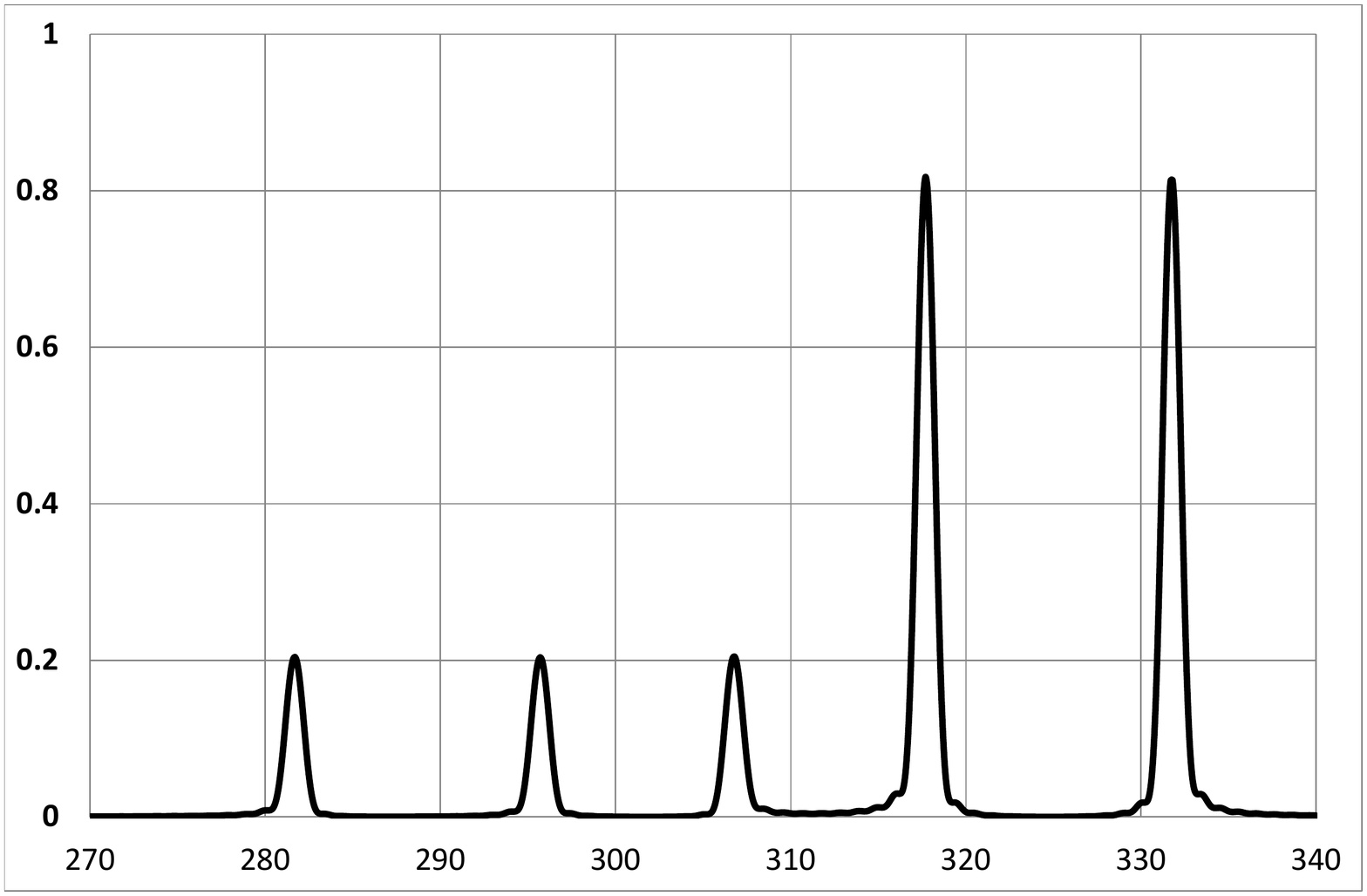}
      \caption{Calculated Power Spectrum of the signal in the setup described in Fig. 2A.
} \label{S2Ai}\end{center}
\end{figure}
\end{document}